# Performance Evaluation of A Proposed Variant of Frequency Count (VFC) List Accessing Algorithm

[1]Rakesh Mohanty, [2]Shiba Prasad Dash, [3]Burle Sharma, [4]Sangita Patel

[1]Department of Computer Science and Engineering, Indian Institute of Technology, Madras, Chennai, India-600036

[2,3]Department of Computer Science and Engineering, Veer Surendra Sai University of Technology, Burla, Sambalpur, Odisha, India

[4]Department of Computer Science Sambalpur University Institute of Information Technology Burla, Sambalpur, Odisha, India

e-mail: rakesh.iitmphd@gmail.com[1], titun.tiki@gmail.com[2], burlesharma@gmail.com[3], sangitapatelsec@gmail.com[4]

*Abstract* - Frequency Count (FC) algorithm is considered as the static optimal algorithm for the list accessing problem. In this paper, we have made a study of FC algorithm and explore its limitation. Using the concept of weak look ahead, we have proposed a novel Variant of Frequency Count (VFC) list accessing algorithm. We have evaluated the performance of FC and our proposed VFC algorithm experimentally using input data set from Calgary Corpus. Our experiments show that for all request sequences and list generated from the above data set VFC performs better than FC.

*Keywords:* Alogrithm, Data Structure, Linked List, Linear Search, List Accessing Problem, Frequency Count Algorithm

## I. INTRODUCTION

An unsorted *linear list* is one of the simplest data structure on which one can perform operations such as *insertion, deletion* and *access*. As access operation is a special case of insertion and deletion operation, we consider only access operation for simplicity. In the *list accessing problem*, a *list* and a *request sequence* are given as input. A *request* is an access operation on any element of the list. A *request sequence* consists of elements of the list with or without repetitions in any order. To process a request from a request sequence, the list has to be traversed linearly from the front until the requested element is found. When an element is accessed, some *access cost* is assigned using a *cost model*. After processing a request, a list may be rearranged by incurring some *reorganization cost*. When a request sequence is processed on a list, a *list accessing algorithm* is used to reorganize the list and to minimize the total access and reorganization cost.

*a) List Accessing Algorithms*
Basically the list accessing algorithms can be classified into two types i.e. *offline* and *online*. In *offline algorithm*, the whole request sequence is completely known beforehand. In case of *online algorithms* the request sequence is either partially known. Here the sequence of requests are served one by one on the fly in the order of their arrival. In comparison to online algorithms, offline algorithms are more powerful. Three well known and widely used primitive list accessing algorithms are Move-To-Front (MTF), Transpose (TRANS) and Frequency-Count (FC). MTF moves the requested element to the front of the list whereas TRANS exchanges the requested element with the element that immediately precedes it. FC maintains a frequency count for each element of the list. After each access of an element, its frequency count is incremented by 1. The elements are rearranged in the list in non-increasing order of frequency count. Most of the list accessing algorithms developed are variants of the above three algorithms. In this work we have made a study of FC algorithm and our objective is to improve the FC algorithm.

*b) Applications and Motivation*
The list accessing problem has been studied as a problem of significant practical interest with many applications. List accessing algorithms are applied while storing data in small dictionaries. Some major applications of list accessing algorithms are in data compression, computing convex hull in computational geometry and resolving collisions in a hash table.

MTF has been proved to be the best performing online algorithm [15] in the literature till date. However not much rigorous study and analysis have been done for the FC algorithm in the literature through FC is considered to be a static optimal algorithm. This motivates us to study the FC algorithm and explore the scope to improve its performance.

*c) Literature Review*
List accessing problem was first studied by McCabe in 1965 [1]. He demonstrated the problem of maintaining a serial file with relocatable records and proposed two algorithms MTF and TRANS. From 1965 to 1985, list accessing problem was studied in [1], [2], [3], [4], [5] with the assumption that the request sequence is generated by a probability distribution. Hester and Hirschberg[6] have presented an extensive survey of average case analysis of list accessing algorithms with some challenging open problems. Sleator and Tarjan[6] have shown the competitiveness of MTF using amortized analysis in the their seminal paper . Albers introduced the concept of strong and weak look ahead in list accessing problem in[7]. Reingold and Westbrook [8] have proposed an optimum off line algorithm for the list accessing problem in 1996. Bacharach and et.al. [9] have provided an extensive theoretical and experimental study of online list accessing algorithm in 2002. A study of list accessing problem with locality of reference was initiated by Angelopoulos in 2008 in which MTF has been proved as the best online algorithm[10]. A comprehensive survey of online list accessing algorithms and associated results are mentioned in [12]. Various results related to FC algorithm have been mentioned in [4], [6] and [9].

*d) Our Contribution*
We have studied the FC algorithm and explore its limitation. In FC algorithm, if the elements having higher frequencies are located towards the end of the list, access cost will increase for subsequent accesses of these elements. This is due to the fact that elements having higher frequencies will not move towards the front until those elements are served in the request sequence and their frequencies exceed the frequency of other elements present in the front of them in the list. To overcome this limitation, we have proposed a new variant of FC algorithm to improve its performance using weak look ahead concept as mentioned in [8]. Our proposed VFC algorithm is tested experimentally using Calgary corpus. We have compared the performance of FC and VFC. VFC performs better than FC as per our experimental results.

*e) Organisation of the Paper*

The paper is organized as follows. Section-I contains the introduction of the list accessing problem. Preliminaries and illustration of FC are presented in Section II. Section III contains our proposed VFC algorithm. Our experimental results have been presented in Section IV. Section V contains concluding remarks.

## II. PRELIMINARIES

A cost model is used to compute the access cost and reorganization cost of a list accessing algorithm. We present two most widely used cost models as follows.

*a) Cost Models*

The two most widely used cost models are *Full Cost Model* and *Partial Cost Model*. In *full cost model*, the cost of accessing an i<sup>th</sup> element in the list from the front is i. In *partial cost model*, the cost of accessing an i<sup>th</sup> element from the front of the list is i-1 because we have to make i-1 comparisons before accessing the i<sup>th</sup> element in the list. After accessing an element in the list, the accessed element can be moved to any position closer to the front of the list without paying any cost using a *free exchange*. But in case of a *paid exchange*, two adjacent elements of the list are interchanged by paying a unit cost

*b) Illustration of FC*

In this section we present an illustration of FC algorithm as shown in Figure 1. Here we have considered 123 as the list and 122323 as the request sequence for our illustration.

## III. OUR PROPOSED ALGORITHM

*a) Uniqueness of Our Approach*

We have considered the concept of weak look ahead for our proposed algorithm. In case of weak look ahead of amount k, the algorithm can foresee next k consecutive requests (may not be distinct) in the request sequence before accessing an element in the list. Here the amount of look ahead is fixed for an algorithm. But in our proposed algorithm, we have taken a variable size look ahead based on some condition. Introduction of variable size look ahead is an attempt to improve the FC algorithm for all request sequences and input lists.

In our proposed algorithm we have considered the full cost model and singly linked list data structure. We have also considered free exchanges while reorganizing the list. In the next section we describe our notations and present the pseudo code for our proposed algorithm as shown in Figure 2.

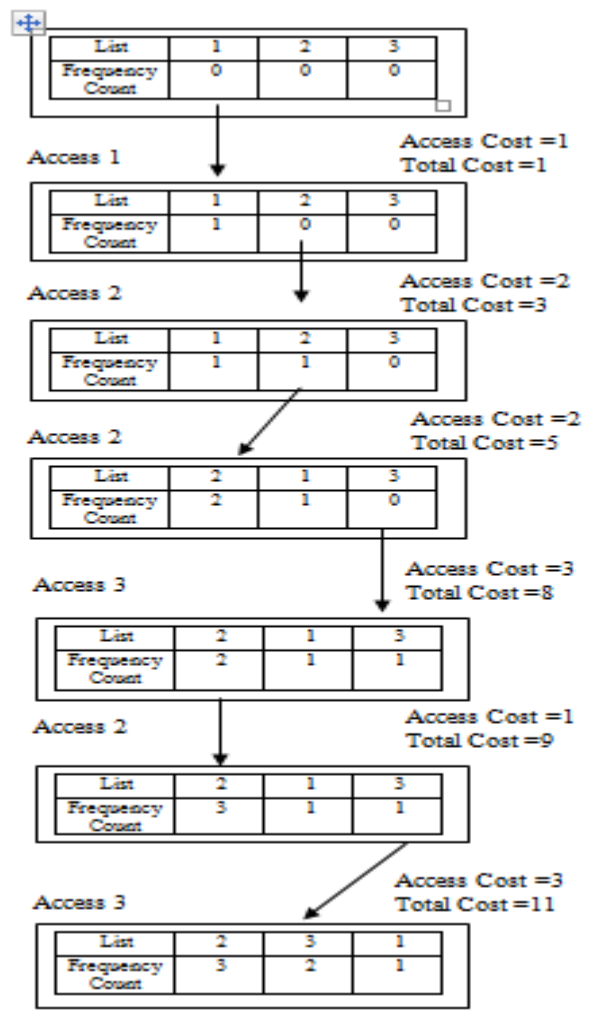

Figure 1. Illustration of FC with an example

*b) Pseudo code*

```
l : linked list
σ : request sequence
l_j : j^th element of the list from the front
F_{l_j} : Current Frequency Count of l_j
σ_i : Accessed Element from σ
F_{σ_i} : Current Frequency Count of σ_i in l
F_h : Frequency Count of head element
F_t : Frequency Count of front element
L_a : Amount of Look ahead
P_{σ_i} : Position of σ_i in l before access
C_a : Access Cost

Procedure Frequency_Count(l, σ_i, F_{σ_i})
For i = 1 to j − 1
{
    If (F_{σ_i} > F_{l_j})
        Reorganise the list so that σ_i
        appears at position i in the linked list
    else if (F_{σ_i} = F_{l_j})
            If (F_{σ_i} > F_{l_{j+1}})
                Reorganise the list so that σ_i
                appears at position i in the linked list
            else
                i + +
    else
        i + +
}
```

```
VFC()
Read σ₁ in σ and Access σ₁ in l;
If (σ₁ is found in l)
{
    C_a = C_a + P_{σ_i}
    F_{σ_i}++
    Call Procedure Frequency_Count(l, σ_i, F_{σ_i})
    F_h = F_t
}
For i = 2 to n
{
    Read σ_i in σ and Access σ_i in l;
    If (F_h ≤ F_{σ_i})
    {
        C_a = C_a + P_{σ_i}
        F_{σ_i}++
        Call Procedure Frequency_Count(l, σ_i, F_{σ_i})
        F_h = F_t
    }
    else
    {
        L_a = |F_a - F_h| + 1
        Take L_a in σ
        If σ_i is present in σ upto L_a
        {
            F_{σ_i} = F_{σ_i} + L_a
            C_a = C_a + P_{σ_i} + L_a - 1
            Call procedure Frequency_Count(l, σ_i, F_{σ_i})
            F_h = F_t
        }
        else
        {
            Call Procedure Frequency_Count(l, σ_i, F_{σ_i})
            F_h = F_t
            C_a = C_a + P_{σ_i}
            F_{σ_i} = F_{σ_i} + 1
        }
    }
}
```

Figure 2. Pseudocode for VFC algorithm

c) Illustration of VFC

Here we have taken a weak look ahead of variable size is not fixed. The amount of look ahead is given by

$$L_a = |F_a - F_h| + 1$$

Suppose the list is 123 and the request sequence is 122333. We initialize the frequency counter of each element of the list to 0 as shown below.

| List | 1 | 2 | 3 |
|---|---|---|---|
| Frequency Count | 0 | 0 | 0 |

| Request Sequence | 1 | 2 | 2 | 3 | 3 | 3 |
|---|---|---|---|---|---|---|

When request 1 is served by VFC the frequency counter of 1 in the list is updated and the configuration of the list is as follows.

| List | 1 | 2 | 3 |
|---|---|---|---|
| Frequency Count | 1 | 0 | 0 |

Now the next element to be accessed is 2 having frequency 0 in the list and the front element 1 of the list having frequency 1. So, we take a look ahead of mod[ (0-1)+1] =2. After accessing two elements in the request sequence, VFC algorithm is executed as mentioned in the pseudo code. Then the configuration of the list along with frequency count is as follows.

| List | 2 | 1 | 3 |
|---|---|---|---|
| Frequency Count | 2 | 1 | 0 |

Doing the same procedure for the element 3 we get the final configuration as

| List | 3 | 2 | 1 |
|---|---|---|---|
| Frequency Count | 3 | 2 | 1 |

So the total access cost of VFC is 1+2+1+3+1+1 =9.

IV. EXPERIMENTAL RESULTS AND ANALYSIS

a) Dataset

We have taken the Calgary Corpus as the data set for our experiment which consists of 14 different files. Out of these 14 files we have considered only 9 files for this experiment. These files represent the information in text form consisting of words. But we have converted all these files by removing all the spaces and return key such that these files consist of only characters.

b) Experiment Performed

Two files are taken as input one for the request sequence and another for the list. A singly linked list is generated for the request sequence and another lined list generated for the list by selecting all the distinct characters. Then these two files are provided as inputs for the two algorithms FC and VFC. The access cost is computed for both the algorithms FC and VFC by taking 9 different files from the Calgary Corpus. We have developed the source code and implemented the above algorithms using C programming in Turbo C++ with Windows XP operating system.

c) Experimental Results

The results of our experiment are presented in Table-I. The input files, size of the request sequence, size of the generated list and corresponding access cost that are computed for FC and VFC algorithms using our data set are shown in Table I.

Table 1 Computation of access cost for FC and VFC

| File Name | Size of Req Seq | Size of List | Access Cost FC | Access Cost VFC |
|---|---|---|---|---|
| Trans | 74476 | 66 | 6501 | 6112 |
| Book1 | 141643 | 72 | 7740 | 7027 |
| NEWS | 310699 | 76 | 8626 | 8139 |
| BIB | 91242 | 77 | 8855 | 8025 |
| PAPER1 | 44309 | 83 | 10292 | 9813 |
| PROGPR | 35398 | 86 | 11051 | 10328 |
| PROGC | 29956 | 89 | 11837 | 11752 |
| GEO | 1345 | 91 | 12335 | 12090 |

Using the data of Table-I, we plot a graph by taking input files in X-axis from which various sizes of list and request sequences are generated. The computed access costs for two algorithms FC and VFC for the generated list and request sequences are represented in Y-axis. The comparison of performances of FC and VFC in terms of access cost has been done and presented in Figure-3.

We have observed that the access cost performance of VFC is better than FC from our experimental results and plotted graph.

V. CONCLUDING REMARKS

Most of the research works carried out on the list accessing algorithms have taken the MTF algorithm for their empirical studies. In our work we have made an analytical study of frequency count list accessing algorithm to explore its limitation. To overcome the limitation we have proposed a novel variant of frequency count algorithm known as VFC using the con-cept of weak look-ahead. Our proposed VFC algorithm is observed to be performing better than FC algorithm for Calgary Corpus data set as evident from our experimental results.

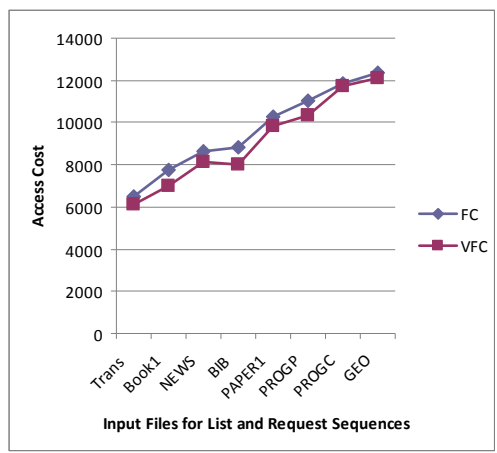

Figure 3. Access Cost Comparison of FC and VFC algorithm